\documentclass{elsart}

\usepackage{epsf}
\usepackage{epsfig}
\usepackage{cite}
\usepackage{graphics}

\newcommand{\dd}{\mbox{d}}
\newcommand{\II}{\mbox{I}}
\begin{document}
\begin{frontmatter}

\title{\Large\bf QED radiative correction to spin-density matrix
elements in exclusive vector meson production}

\author{I. Akushevich and P. Kuzhir}
\address{National Center of  Particle and  High Energy Physics,
Bogdanovich str. 153, 220040 Minsk, Belarus}

\begin{abstract} QED radiative effects are considered in the case of
  measurement of spin-density matrix elements of diffractive
  $\rho$-meson electroproduction. Large radiative correction
for  $r^5_{00}$ is found in the kinematics of collider experiments at HERA.
\end{abstract}

\end{frontmatter}

The extension of the kinematical region of lepton-nucleon deep
inelastic scattering to the domain of the diffractive processes
provides hadronic nature of the photon to be studied along with
the nucleon structure. Particular emphasis has been placed on the
case of exclusive vector meson production
\begin{equation}
e(k_1) + p(p) \rightarrow e(k_2) + p(p') + \rho (p_V), \;\;\;
\rho \rightarrow \pi ^+ (p_+) + \pi ^- (p_-).
\label{eq1}
\end{equation}
The reason is
the process (\ref{eq1}) can be viewed as an
off-diagonal Compton scattering analytically continued in the
virtuality of the photon $\gamma ^*$ to the vector meson mass
$\gamma ^* p \rightarrow V p$ and
gives access to the whole set of the corresponding helicity
amplitudes.

The process (\ref{eq1}) is analyzed experimentally by
means of
spin-density matrix  elements. When measured, they
give an indication of vector meson internal constituents motion
and its spin-angular structure.
The angular distribution of unpolarized vector meson decay is
parameterized by fifteen matrix elements
$r_{ij}^{\alpha }, r_{ij}^{\alpha \beta }$.
For a long time it was believed, that their
behavior complies with
the s-channel helicity conservation (SCHC)  hypothesis, which means
that the helicity of the virtual photon is conserved in the
s-channel process $\gamma ^* p \rightarrow \rho p$. In this case
ten matrix elements (which corresponds to the
case when
photon and vector meson have different helicities) are equal to
zero. But in the recent measurements
$r^5_{00}$ has been observed to be non zero \cite{ZEUS1,H1a,ZEUS2}, what
has been considered as
an indication to SCHC violation.

The procedure of the experimental data analysis is based on the
correlation of the lepton scattering, vector meson production
and decay planes, which are
affected by the radiative corrections (RC). Hence it is topical
to look at whether the measured $r^5_{00}$ can, at least partly,
be
the result that RC coming from
non-observed QED effects and real
photon emission was underestimated.
In any case in order to
make the data processing of the corresponding experiments \cite{ZEUS1,H1a,ZEUS2}
to  be  consistent, RC should be taken into
account.

Following the analysis of K.Schilling and G.Wolf 
\cite{SW} one reconstructs
$r_{ij}^{\alpha }, r_{ij}^{\alpha \beta }$
through
vector meson decay angular distribution
$W \bigl   ( \cos \theta , \phi , \Phi  \bigr )$  and
the weight coefficients $F_{ij} \bigl   ( \cos \theta , \phi ,
\Phi
\bigr
)$ (see Appendix C of \cite{SW}), the observed
matrix elements
$r_{ij}^{obs}$
 can be written as:
\begin{equation}
r_{ij}^{obs}= \frac {\int\limits _0 ^{2\pi } \frac {\dd\Phi }{2\pi
}
\int\limits
_{-1}^{+1} \dd \cos \theta  \int\limits _0 ^{2\pi } \dd\phi \;\;
W \bigl   ( \cos \theta , \phi , \Phi  \bigr )
F_{ij} \bigl   ( \cos \theta , \phi , \Phi  \bigr ) (1+\delta )
 }
{\int\limits _0 ^{2\pi } \frac {\dd\Phi }{2\pi } \int\limits
_{-1}^{+1} \dd \cos \theta  \int\limits _0 ^{2\pi } \dd\phi \;\;
W \bigl   ( \cos \theta , \phi , \Phi  \bigr )
(1+\delta ) }.
\label{eq4}
\end{equation}
Here
$\Phi $ is the angle between the lepton scattering
plane and $\rho-$production plane,
$\phi $ is the angle between $\rho-$decay and production plane,
$\theta $ is the polar angle
of the direction of flight of the positive decay pion.
$\delta $ is RC obtained as the ratio of next
to the lowest order cross section of the process (\ref{eq1}) to the Born
cross section.

The RC in the case that
vector  meson  in  the  final  state is considered as a stable
particle
 was calculated in \cite{Ak} and can be
presented in the form
\begin{equation}
\delta = e^{\delta _{inf}} (1+\delta _{VR}+\delta _{vac}) + \frac
{\sigma _F} {\sigma _0}.
\label{eqrc}
\end{equation}
Recall that $\sigma _0$ is the Born cross section,
$\delta _{vac}$ comes from the effects of vacuum polarization by
leptons and hadrons, the sum of $\delta _{VR}$ and $\delta _{inf}$
origins from contributions of vertex function and soft
photon emission (the exponent is due to the multiple soft photon
radiation), and
$\sigma _F$ appears for
the hard photon emission.

Let us discuss the angular dependence of RC (\ref{eqrc}). We
remind that the angle $\Phi $ is described by the vector meson
momentum and nonmeasured vector $\vec q$ (which is determined by
measured momenta of initial and scattered leptons). It is clear
that the radiation of unobserved real photon changes the vector
$\vec q$ into $\vec q - \vec k$ ($\vec k$ is the real photon
momentum) leading in fact to
reorientation of production and scattering planes and, therefore
the hard photon contribution $\sigma _F$ to RC can be
significantly dependent on $\Phi $. The quantities $\delta _{VR}$
and $\delta _{inf}$ depend only weakly on $\Phi $. We note that
this dependence has kinematical origin. It means that if the
integration
region over photon variable is divided into soft
and hard parts\footnote{In this case we would come to
the formulae analogous to ones of traditional
approach of Mo and Tsai \cite{motsai} for deep inelastic
scattering.},
the splitting parameter could be chosen in such a way that
mentioned
dependence would be completely reduced.

    Strictly  speaking,  $\delta  $  was  found in \cite{Ak} as the
ratio of four-fold cross sections $d^4 \sigma / dx dy dt d\Phi
$.
It can be shown, however, that in our case, when  $\rho
-$meson decays into $\pi _+ \pi  _-$ and RC is denoted by  a ratio
of  seven-fold  cross  sections,  the  results  of \cite{Ak} can be
applied unchanged.  All one has to do is to show that momenta  of
$\pi -$mesons are appear in  the correction (3) only as
$p_+ + p_-=p_V$, but not separately. In this case there would be no any
scalar
products of the four momenta of pions
 which can produce dependence on $\cos \theta$ and $\phi$.

Really, $\delta _{vac}$ and infrared finite part of vertex function
contributed to $\delta_{VR}$
are determined by $Q^2$ only. Other contributions to $\delta 's$
in
(\ref{eqrc}) being the result of
the infrared divergence cancellation
 can depend along
with the kinematical variables $Q^2, W^2, t$ also on scalar
products of  vector $\Lambda $
($\Lambda  =p'+k=k_1-k_2-p_+-p_-$ is the four-momentum
of the system of unobserved particles): $\Lambda k_1, \;\;
\Lambda k_2, \;\; \Lambda ^2$, and  therefore only on $\Phi  $.
The real photon phase space is also specified by $\Lambda $, but
not $p_+$ or $p_-$. And at last if the natural
assumption
\cite{Rys}
has been done  that all structure  functions excepting $\sigma
_L$,
$\sigma _T$ vanish in the hadronic tensor, the hard photon contribution
$\sigma
^F$ to RC is also found to be free of
$\cos \theta$, $\phi - $dependence.
As a result, RC
depends only on $\Phi $
and is consequently reduced to one calculated in
\cite{Ak}.

Then by
definition (\ref{eq4}) the QED corrections ($\Delta r=r_{obs}-r_{Born}$)
to  the matrix elements are
\begin{eqnarray}\label{main}
&\Delta {\rm r}^{04}_{00}&=\;-\epsilon \II_2 {\rm r} ^1_{00} + a \II_1 r^5
_{00}
,\nonumber\\[0.3cm]
&\Delta{\rm Re\;r}^{04}_{10}&=\; - \epsilon \II_2 {\rm Re\;r}^{1}_{10}
+ a\II_1 {\rm Re\;r}_{10}^5
,\nonumber\\[0.3cm]
&\Delta{\rm r}^{04}_{1-1}&=\;
-\epsilon \II_{2} {\rm r}^{1}_{1-1} + a \II_1 {\rm r}_{1-1}^5
,\nonumber\\[0.3cm]
&\Delta{\rm r}^{1}_{00}&=\; \frac{1}{\epsilon}
\bigl[
-2 \II_{2}
 {\rm r}^{04}_{00}+\epsilon \II_{4}
 {\rm r}^{1}_{00}-a ( \II_{1}+\II_{3})
{\rm r}^{5}_{00} \bigr]
,\nonumber\\[0.3cm]
&\Delta{\rm r}^{1}_{11}&=\;{1 \over \epsilon}\bigl[ \II_{2}
({\rm r}^{04}_{00}-1)+\epsilon
\II_{4} {\rm r}^{1}_{11}
- a (\II_{1}+\II_{3}) {\rm r}^{5}_{11}
\bigr]
,\nonumber\\[0.3cm]
&\Delta{\rm Re\;r}^{1}_{10}&=\;{1 \over \epsilon}\bigl[-2
\II_{2}
{\rm Re\;r}^{04}_{10}+\epsilon \II_{4}
{\rm Re\;r}^{1}_{10}-a (\II_{1}+\II_{3}) {\rm Re\;r}_{10}^5
\bigr]
,\nonumber\\[0.3cm]
&\Delta{\rm r}^{1}_{1-1}&=\;{1 \over \epsilon}\bigl[-2
\II_{2}
{\rm r}^{04}_{1-1}+\epsilon \II_{4}
{\rm r}^{1}_{1-1}-a (\II_{1}+\II_{3}) {\rm r}_{1-1}^5
\bigr]
,\nonumber\\[0.3cm]
&\Delta{\rm Im\;r}^{2}_{10}&\;=-\II_4 {\rm Im\;r}^{2}_{10}+
{a\over \epsilon} (\II_1+\II_3) {\rm Im\;r}^{6}_{10}
,\nonumber\\[0.3cm]
&\Delta{\rm Im\;r}^{2}_{1-1}&=\;-\II_4 {\rm Im\;r}^{2}_{1-1}
+ {a\over \epsilon} (\II_1+\II_3) {\rm Im\;r}^{6}_{1-1}
,\nonumber\\[0.3cm]
&\Delta{\rm r}^{5}_{00}&=\;{1 \over a}\bigl[
2 \II_{1} {\rm r}^{04}_{00}+a \II_{2} {\rm
r}^{5}_{00}-\epsilon (\II_{1}+\II_{3}) {\rm r}^{1}_{00}
\bigr]
,\nonumber\\[0.3cm]
&\Delta{\rm r}^{5}_{11}&=\;{1 \over a}\bigl[\II_1(1-{\rm
r}^{04}_{00})
+a\II_2 {\rm r}^{5}_{11} - \epsilon (\II_1+\II_3) {\rm r}^{1}_{11}
\bigr]
,\nonumber\\[0.3cm]
&\Delta{\rm Re\;r}^{5}_{10}&=\;{1\over a}\bigl[ \II_1 {\rm
Re\;r}^{04}_{10}
+a \II_{2} {\rm Re\;r}_{10}^5
- \epsilon  (\II_{1}+\II_{3}) {\rm Re\;r}^{1}_{10}
\bigr]
,\nonumber\\[0.3cm]
&\Delta{\rm r}^{5}_{1-1}&=\;{1\over a}\bigl[\II_1 {\rm r}^{04}_{1-1}
+a \II_{2} {\rm r}_{1-1}^5
-\epsilon (\II_{1}+\II_{3}) {\rm r}^{1}_{1-1}
\bigr]
,\nonumber\\[0.3cm]
&\Delta{\rm Im\;r}^{6}_{10}&=\;
-\II_{2} {\rm Im\;r}^{6}_{10}+\frac {\epsilon}{a} (\II_1+\II_3) {\rm
Im\;r}^{2}_{10}
,\nonumber\\[0.3cm]
&\Delta{\rm Im\;r}^{6}_{1-1}&\;=
-\II_{2} {\rm Im\;r}^{6}_{1-1}+\frac {\epsilon}{a} (\II_1+\II_3) {\rm
Im\;r}^{2}_{1-1}.
\end{eqnarray}
Polarization parameter of the virtual photon density matrix
$\epsilon = \frac {1-y}{1-y -y^2/2}$
is close to 1 at HERA kinematics, $a=\sqrt {2\epsilon
(1+\epsilon)}$,
\begin{equation}
\II_{n}=\frac {\int\limits _0^{2\pi } \frac{\dd\Phi}{2\pi} \; \cos n\Phi
\; \delta
(\Phi )}
{\int\limits _0^{2\pi } \frac{\dd\Phi}{2\pi}\; (1+\delta
(\Phi ))}, \;\;\; n=0,...,4.
\end{equation}
    Thus the absolute radiative correction to spin-density  matrix
elements  is  linear  in  the  lowest  order of $r_{ij}^{\alpha   },
r_{ij}^{\alpha \beta }$, and the dependence on
$\delta=\delta(\Phi)$
is included in the coefficients $I_n$.

\begin{figure}[t]
\includegraphics[height=7cm]{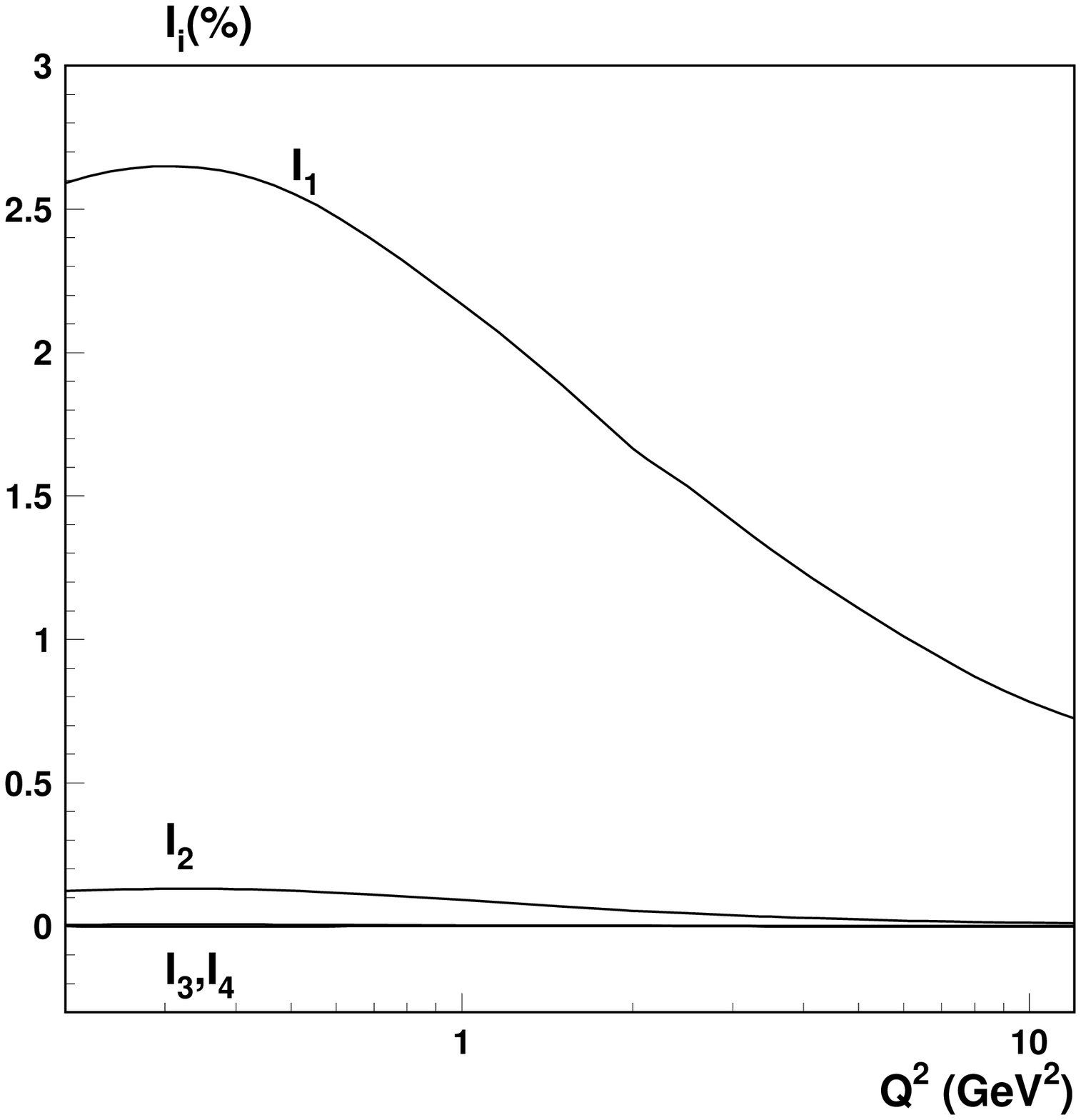}
\hfill
\includegraphics[height=7cm]{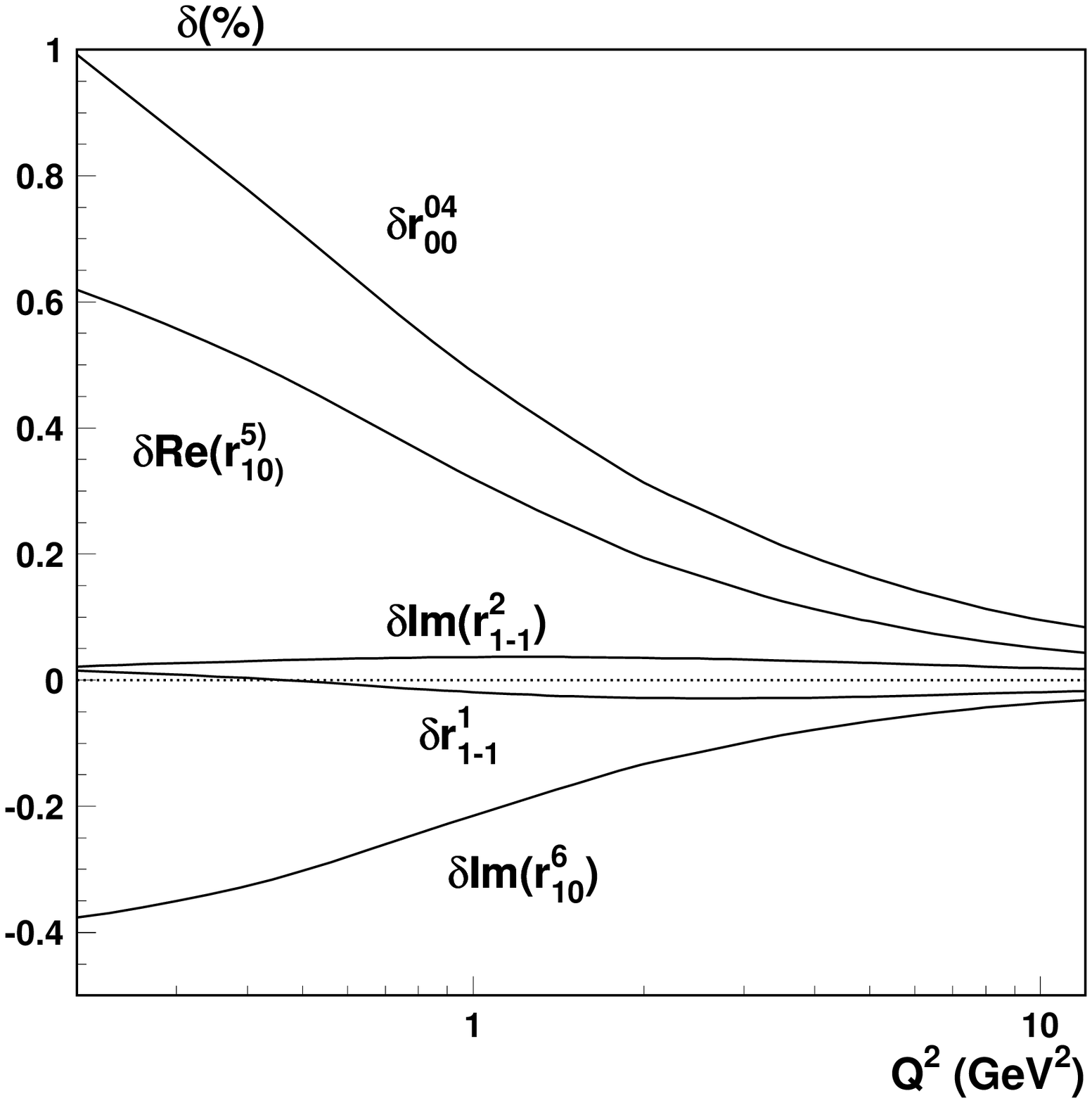}
\\
\parbox[t]{0.46\textwidth}{\caption{The  dependence  of  $I_n,  n=1...4$  on  $Q^2$  under  the
kinematical conditions of H1/ZEUS  experiments:
$\sqrt s = 300$ GeV, $W=75$ GeV.}\label{fig2}}
\hfill
\parbox[t]{0.46\textwidth}{\caption{The relative RC $\delta  r$ to non-zeroth in accordance  with
SCHC matrix elements under  the kinematical conditions of  H1/ZEUS
experiments. }\label{fig4}}
\end{figure}

It is clear, that Born matrix elements can be easily extracted
from
formulae (\ref{main}) without using any model for ${\rm r'}s$.
For realistic radiative correction procedure
a system of equations (\ref{main}) can be solved by
the simplest and traditional way: to perform
the iteration procedure, where extracted matrix element at the $n$
step is calculated via ${\rm r'}s$ estimated at the $n-1$ step as
\begin{equation}
{\rm r}_{ext}^{(n)}={\rm r}_{obs}-\Delta {\rm r}({\rm
r}_{i\;ext}^{(n-1)}).
\end{equation}

Note that the value of RC
$\Delta r $ is expected to
be small in respect to correction to the cross section.  The
reason is
only  contributions  of  higher  harmonics  $I_n$ survive, but the
large contribution of $I_0$ vanishes.  The  quantities
$I_{1,2,3,4}$ are shown in Figure  \ref{fig2}.  $I_1$ is of  order
1-2\%,  $I_2$  is  less  then  1\% while $I_{3,4}$ are practically
negligible in the considered kinematical region.
It follows that
only those $\Delta{\rm r}$ would
be significantly different from zero, which are proportional
to non-vanished matrix elements with relatively large coefficient
$I_1$.

In order to treat the radiative effects numerically
the model \cite{NNN} reasonably reproduced
experimental data has been attracted.
We estimate the
relative RC $\delta{\rm r} = \Delta{\rm r} / {\rm r}$ for those
matrix elements, which should according to SCHC be non-zeroth.
 We found (see Figure 2) that in kinematics of experiments at HERA
$\delta{\rm r}$ do not exceed 1\%.
Let us stress, that if SCHC is true,
$\delta {\rm r} ^{04}_{00}$ will be identically equal to zero,
since it is proportional
to zeroth (by SCHC) matrix elements ${\rm r} ^{1}_{00}$, ${\rm r}
^{5}_{00}$.

\begin{figure}[t]
\includegraphics[height=7cm]{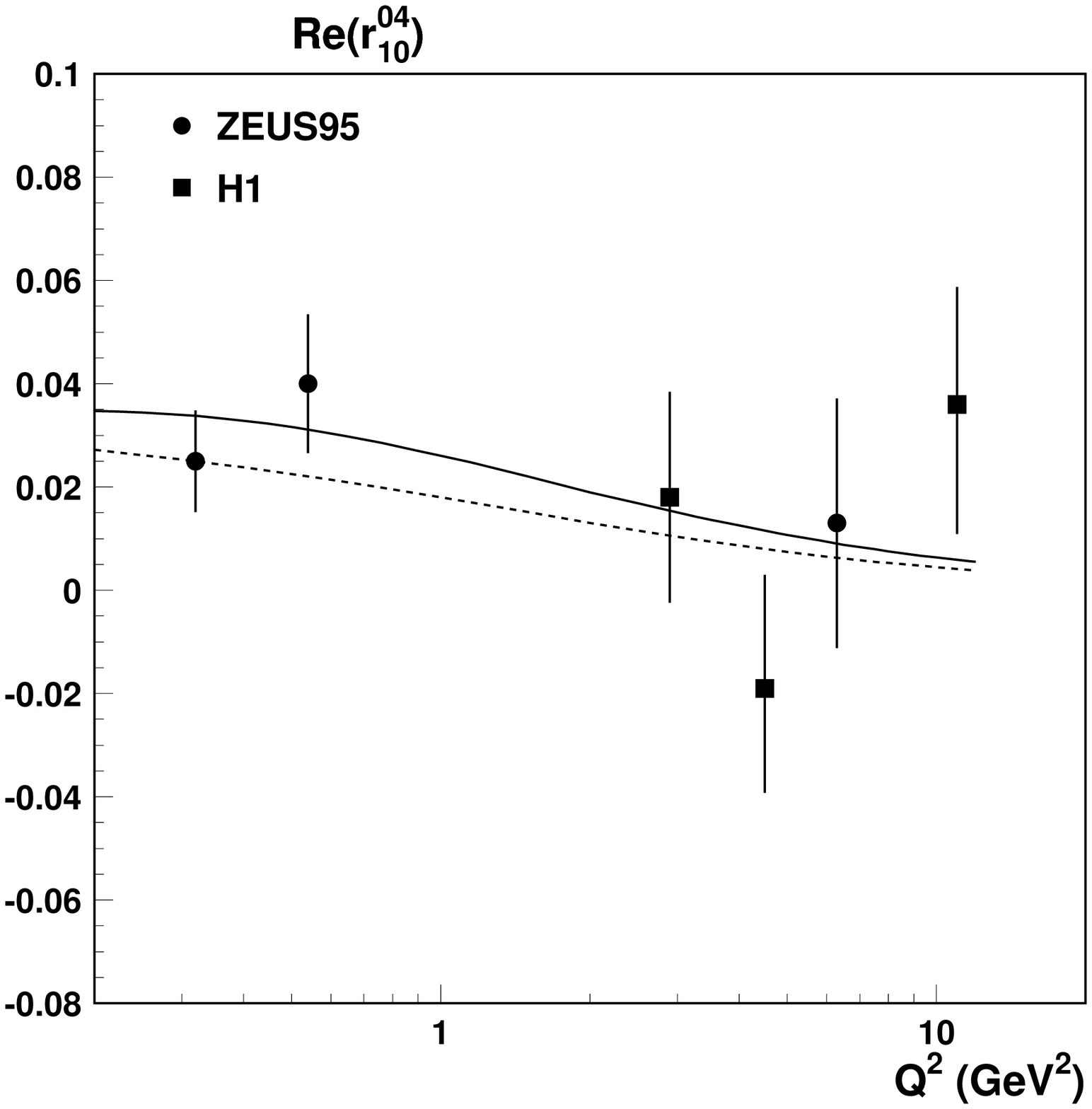}
\hfill
\includegraphics[height=7cm]{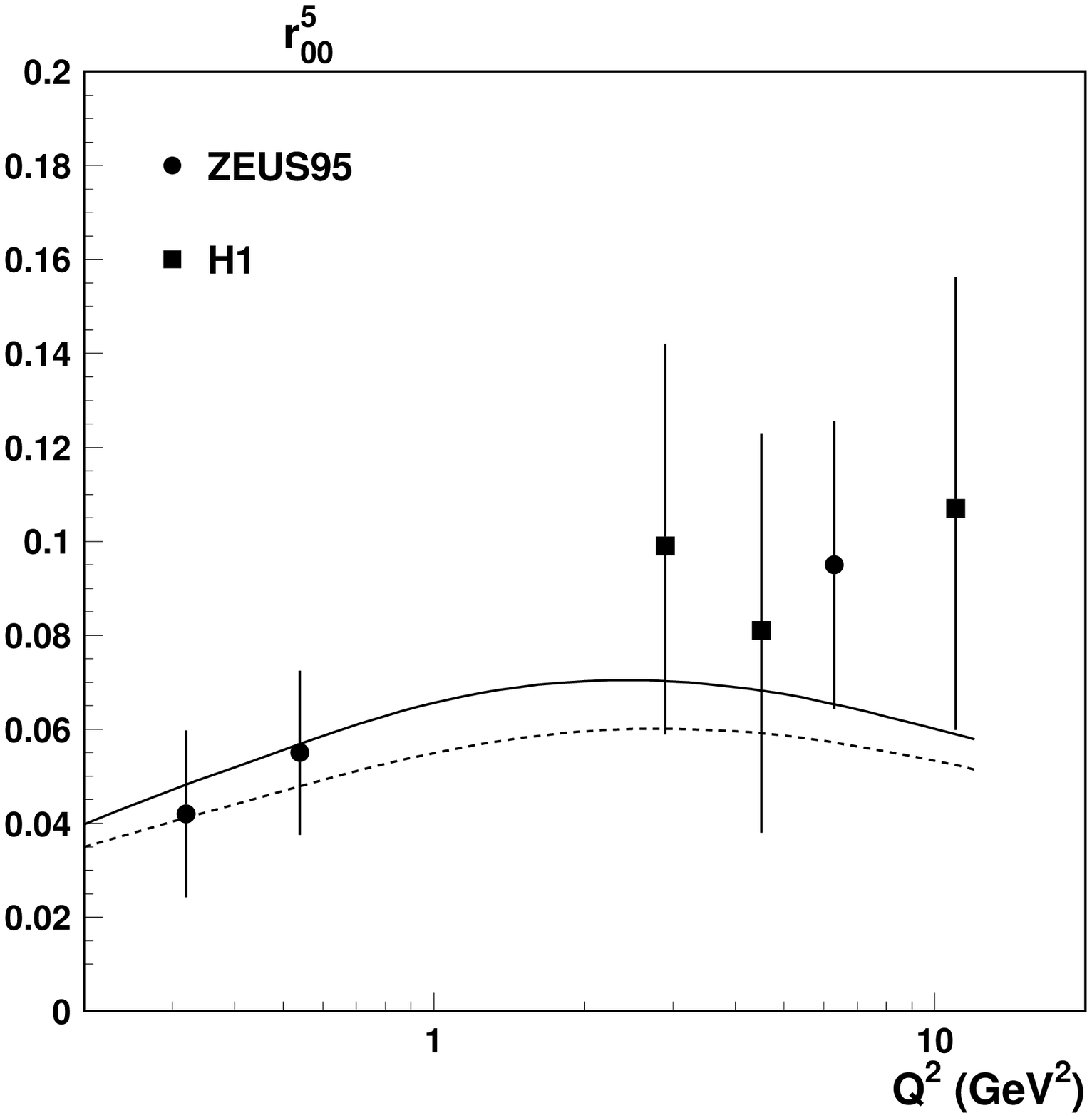}
\caption{The dependence of  Born (dashed line) and
radiative corrected
(solid  line) spin-density  matrix elements  on $Q^2$ under
the   kinematical   conditions   of   H1/ZEUS  experiments.
 }
\label{fig1}
\end{figure}

For the majority of the matrix elements vanishing in the SCHC
limit, radiative corrections turn out to be not greater then 1\%.
  However there are two of them,  ${\rm Re\;r}^{04}_{10}$ and
${\rm r}^{5}_{00}$, which RC
 appears to be
substantial (see Figure \ref{fig1}). One can see, that
corrections  $\Delta {\rm Re\;r}^{04}_{10}$ and $\Delta {\rm
r}^{5}_{00}$ may reach $\sim $ 20\%.


The last result is
interesting from point
of view of the found SCHC violation: the
radiative
correction procedure reduces the observed effect.

\begin{figure}[t]
\centering
\includegraphics[height=7cm]{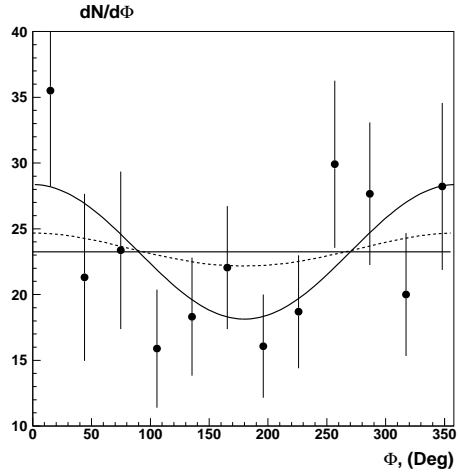}
\caption{
Vector meson decay $\Phi- $distribution: SCHC curve (straight
line), experimental curve
\cite{H1a}  (solid  line)  and  radiative corrected theoretical
curve
(dashed line). }
\label{fig5}
\end{figure}

As an illustration let us follow the origin of
the radiative effect within the
experimental data processing.
${\rm r}^{5}_{00}$ is defined experimentally by
fitting of vector meson decay $\Phi$ distribution  ($W
\bigl   ( \cos \theta , \phi , \Phi  \bigr )$ integrated over
$\cos \theta , \phi $)
\begin{equation}
W(\Phi ) \sim 1-\epsilon \cos 2\Phi (2{\rm r}^1_{11} + {\rm
r}^1_{00})
+a
\cos
\Phi (2 {\rm r}^5_{11}+{\rm r}^5_{00}).
\
\end{equation}
Based on SCHC hypothesis
this distribution would be
flat, what corresponds to
zeroth
${\rm
  r}^{5}_{00}$.
But this is
true only for matrix elements in the lowest order of QED.
It can be seen (Figure \ref{fig5}, see also \cite{Ak2}),
that the theoretical radiative corrected $\Phi-$distribution
deviates from flat, and
has the form similar to the experimental
distribution. Thus the observed effect  comes not only from
SCHC violation but due to radiative corrections as well.

It follows that
if RC procedure is
included in the data processing, it would lead (as one can see
from Figures \ref{fig1} and \ref{fig5}) to
reducing of the found effect of SCHC violation almost on 20\%.

\bigskip

We are grateful to  A.B.Borissov, N.N.Nikolaev, A.S.Proskuryakov,
who took their time to read this paper attentively and discuss it
with us. Their advices and suggestions were very helpful for us.

{}

\end{document}